# The Research Space: using career paths to predict the evolution of the research output of individuals, institutions, and nations

Miguel R. Guevara[1,2,4], Dominik Hartmann[1,3], Manuel Aristarán[1],
Marcelo Mendoza[4], and César A. Hidalgo[1*]

[1]Macro Connections, The MIT Media Lab, Massachusetts Institute of Technology, Cambridge, MA,
[2]Department of Computer Science, Universidad de Playa Ancha, Valparaiso, Chile,
[3]Chair for Economics of Innovation, University of Hohenheim, Germany,
[4]Department of Informatics, Universidad Técnica Federico Santa María, Santiago, Chile

**Abstract:**

In recent years scholars have built maps of science by connecting the academic fields that cite each other, are cited together, or that cite a similar literature. But since scholars cannot always publish in the fields they cite, or that cite them, these science maps are only rough proxies for the potential of a scholar, organization, or country, to enter a new academic field. Here we use a large dataset of scholarly publications disambiguated at the individual level to create a map of science—or *research space*—where links connect pairs of fields based on the probability that an individual has published in both of them. We find that the research space is a significantly more accurate predictor of the fields that individuals and organizations will enter in the future than citation based science maps. At the country level, however, the research space and citations based science maps are equally accurate. These findings show that data on career trajectories—the set of fields that individuals have previously published in—provide more accurate predictors of future research output for more focalized units—such as individuals or organizations—than citation based science maps.

# Introduction

While most scientists are trained in one specialized academic field, their scholarly contributions usually involve multiple fields. In fact, 99.8% of the 215,390 scholars that had a Google Scholar profile by May 24, 2014, and that received citations in at least ten different papers, had published in two or more academic fields (with fields defined according to the 308 categories in the SCImago classification of journals from Scopus). But trans-disciplinary efforts are not constrained to pairs of disciplines. In fact, 99.2% of these scholars had also published in three or more fields, and 97.5% of them in four or more. These numbers show that the work of most scholars is not constrained to a single academic discipline, but often spans at least a few of them.

But while most scholars do not publish in a single discipline, their contributions are nevertheless confined to a small set of highly related fields. Consider, for instance, the 24,125 scholars in our dataset (see Data and Methods) that have published at least two papers in "Molecular Biology." 46.6% of these scholars also had published in "Clinical Biochemistry," but only 0.95% of them also published in "Economics and Econometrics." Since the total number of scholars with at least two papers in "Clinical Biochemistry" (11,110) is similar to the number of scholars with at least two papers in "Economics and Econometrics" (10,479), the larger overlap of the first pair vis-à-vis the second, tells us that "Molecular Biology" is more related to "Clinical Biochemistry" than to "Economics and Econometrics."

But the structure of these academic overlaps is not theoretically surprising. Scholars are often trained in narrowly defined academic disciplines, and they spent most of their careers in relatively homogenous academic departments. This homogeneity in training also leads to relatively high levels of homogeneity in their social and professional networks. An illustration of this social homogeneity is the large number of marriages among scientists—a proxy for strong links in a social network. Marriages among scientists go as high as 56% for women scientists in their first marriage, and 63% for women scientist in their second marriage (compared to 14% and 32% for males) [1].

Among women in the first marriage, 36% marry a scholar within the same field. Thus, the professional and social institutions where scholars are embedded [2] reduce the opportunity for scholars to develop the contacts, or skills; they need to enter "distant" academic fields. As a result, the diversification paths followed by individuals, organizations, and countries, are constrained by the homogeneity of the social networks of scholars and their professional institutions. These various constraints should be reflected in the structure of the network connecting related academic fields.

But the prevalence of researchers publishing in multiple academic fields is good news for those looking to either predict the evolution of research production, or evaluate the potential of an organization to enter a particular academic field. In fact, the overlapping participation of scholars in related disciplines tells us about the possible career paths of scholars. Moreover, since research organizations, and national research efforts, are composed of networks of scholars, the network of related academic disciplines should be predictive of the probability that a country or organization will enter a new academic field.

Here we leverage information on the observed career paths of more than two hundred thousand scholars to introduce the *research space*, a map connecting pairs of fields based on the probability that an author has published in both of them. We argue that this map captures implicit information about the skills, social networks, and institutions constraining the movement of scholars into different academic disciplines. We validate the predictive superiority of the research space by using Response Operator Characteristic curves (ROC curves) and show that the research space is a more accurate predictor of the future presence of an individual or organization in an academic field than citation based or knowledge flow science maps.

**Mapping Science through Knowledge Flows and Career Paths**

In recent decades bibliometricians, information scientists, sociologists, physicists, and computer scientists, have created maps of science connecting fields that either cite each other, or that cite similar literature [3–5]. These citation based maps of science, or knowledge flow maps, tell us if the knowledge developed in one field is used to produce knowledge in other fields. Ultimately, these maps help us categorize science and understand the trans-disciplinary impact of scholarly work.

Most knowledge flow science maps use one of three methods: co-citation, direct citations, or bibliographic coupling. Co-citation networks [4,6–8] connect academic disciplines by looking at the reference section of a paper and connecting the areas of the papers that appear in the same list of references (i.e. they connect papers A and B, if paper C cites both of them) (Figure 1 a). Direct citation networks, on the other hand, [4,5,9] connect academic disciplines when a paper from one discipline cites a paper from another discipline (Figure 1 b). Direct citation networks includes both, networks where scholars differentiate the source and target fields, and un-directed networks, where information on what field is citing, and what field is cited, is disregarded. Finally, bibliographic coupling networks [3,4], connect pairs of disciplines when papers from different fields cite the same other papers (Figure 1 c).

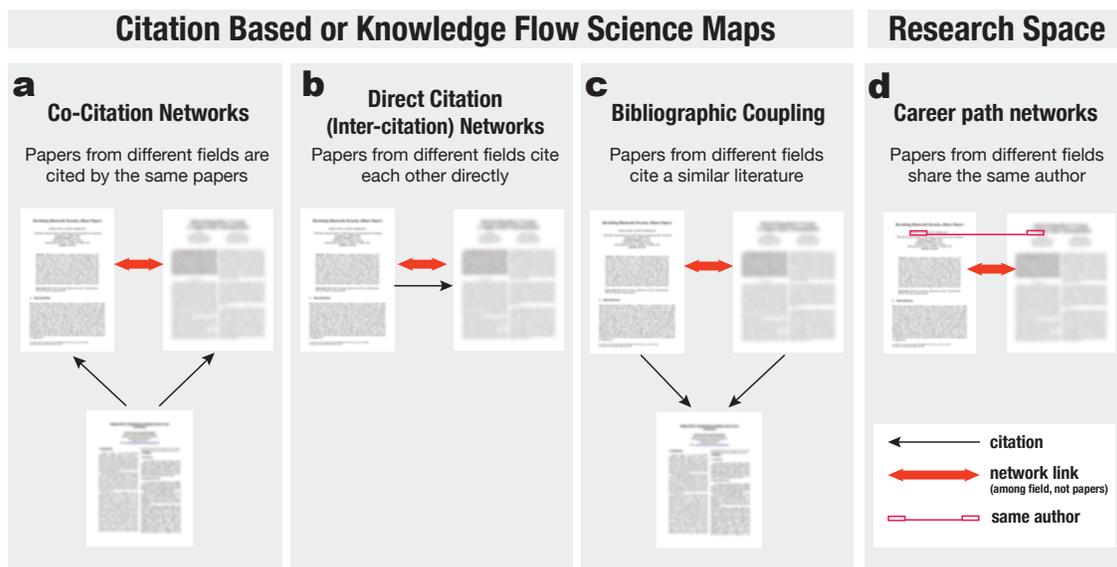

**Figure 1**: **Methods used to create science maps.** Citation based or Knowledge flow Science maps include: Co-citation networks, that connect the academic fields of papers that appear in the same reference sections; Direct citations networks, that connect fields when papers from these fields cite each other; and Bibliographic coupling networks, that connect fields that cite a similar literature. The Research Space is not based on citations and connects fields when researchers are likely to have published in both of them.

Beyond citation-based maps, scholars have also used online searchers to connect academic disciplines. The Clickstream Science Map by [10] connects academic disciplines based on the probability that a scholar who searched for a paper from one field, also searched for a paper from another field. In spirit, the clickstream map is similar to the networks created from co-citations or bibliographic coupling because it also focuses on knowledge flows. Yet since online searches are a more common expression of interest in a topic than a formal citation (the latter requires the costly process of publication), efforts like clickstream help leverage new datasets that are more dynamic than those based on citations.

But what are these science maps used for? A common use of knowledge flow maps is to categorize knowledge. The idea of knowledge categorization has a long tradition in bibliometrics, going back at least to the work of Paul Otlet, the creator of the Universal Decimal Classification, and Ramon Llull, the creator of the XIV century science tree. This idea, however, continues to be influential in recent projects, such as the consensus Map of Science [11] or the UCSD Science Map and Classification System [3]. The UCSD science map has been used to construct a classification of 554 research areas that some university libraries now use to understand the research production of their scholars. Another example of the use of science maps includes the cross-citation maps of Leydesdorff and Rafols [5], who overlaid the research structure of universities [12] to contextualize a university's research output.

Science maps can also be powerful policy instruments. In a world where research budgets are constrained, and the probability of succeeding in a field is uncertain, science promotion agencies (like the N.S.F. in the U.S., the F.A.P.s in Brazil, or the C.N.R.S. in France) need to decide the amount of funds they will allocate to each field, including those where a country or institution may not have a presence and the probability of success is uncertain. Science maps can help estimate a field's strategic value, by helping administrators estimate the probability of success, and therefore the cost, of venturing into a new research area.

But research fields are not only connected by the knowledge flows that are expressed in citations. Since scholars around the world participate in multiple fields, information about the career trajectories of scholars (Figure 1 d) represents a viable alternative to knowledge flow maps. In fact, career trajectories have been used to create predictive maps in other areas of research. For instance, labor flows among industries have been used to study the stability of industrial clusters [13], and the labor mobility of displaced workers [14]. Labor flows among occupation have also been used to create online tools that help visualize the possible career paths of workers or the industrial evolution of cities [15].

Here, we use the career trajectories of hundreds of thousands of scholars to create a map of science—or research space—to predict the future research output of countries, organizations, and individuals. We find that for the most disaggregate units (individuals and organizations) the research space is a more accurate predictor of the development of future research areas than knowledge flow based science maps.

## Data & Methods

**Data**

Research maps where links connect areas sharing authors are uncommon because most datasets on research production are not properly disambiguated at the author level (i.e. these datasets lack the ability to distinguish among authors with similar names). Here, we solve the disambiguation problem by looking only at data from authors who have created a profile in Google Scholar. We note that the Google Scholar dataset is not free of biases, as the adoption of Google Scholar is not uniform across academic fields, or age groups. So we interpret our results in the narrow context of the data used to produce them. These results are applicable only to the career trajectories that are observable in Google Scholar.

We filter this dataset by focusing only on scholars with less than fifty publications in each year, because those with more than fifty publications tend to have many publications that are miss-assigned and are not theirs (see supplementary material for more details). Our filtered dataset contains 319,049 authors who have authored a total of 4,745,774 publications indexed in 16,873 journals and proceedings between 1971 and 2014 (we note that in the introduction we have a smaller number of authors because there we considered only authors with at least ten papers that have received one citation).

We assign each publication to a research category based on the journal in which it was published using Scopus classification system provided by SCImago that includes 27 main areas of knowledge that are subdivided into 308 fine grained categories. In our dataset we use only the 2 categories for which at least one paper was found (For a complete list of categories see supplementary material).

We also aggregate the author level data by identifying the organization (i.e. the university or research institution) and country where the scholar participates in. We first identify organizations by matching the verified email provided in the Google Scholar profile of the author, and then, assign organizations to countries according to the list of institutions provided by the Webometrics Ranking of World Universities [16].

For comparisons we download the UCSD science map [3], which is a citation based science map based on bibliographic coupling (Figure 1 c) available for download at: http://sci.cns.iu.edu/ucsdmap/. When comparing with the UCSD science map we transform all of our papers to their classification, since in the same website, a one-way mapping from journals to their classification was available.

**Constructing The Research Space**

We begin the construction of our research space by defining the presence of a scientist $s$ in academic field $f$. We define the presence of a scientist $s$ in a field $f$ at time $T$ by taking the sum of the papers produced by scientist $s$ in academic field $f$ before time $T$, normalized by the number of co-authors she had on each paper $p$ denoted by variable $n_p$ and the number of fields of the journal where the paper was published $m_p$ (since a single paper can be assigned to multiple categories depending on the journal). Formally we define the matrix $X_{sf}(T)$ as the summation over all papers $p(s,f,T)$ produced by scientist $s$ in field $f$ before time $T$ as:

$$X_{sf}(T) = \sum_{p(s,f,T)} \frac{1}{n_{p(s,f,T)} m_{p(s,f,T)}}$$

$X_{sf}(T)$ is an indicator of the presence of a scientist in a field that controls for the number of co-authors with which a scientists has published and the number of fields in which a journal is classified. We then discretize $X_{sf}(T)$ to remove scientists that have produced only a marginal contribution to field $f$ (scientists that have only produced a small anecdotal participation in field $f$ in an effort with many co-authors). We remove marginal contributions by creating the matrix $P_{sf}(T)$, which is equal to one if the output $X_{sf}(T)$ of scientist $s$ in field $f$ is larger than 0.1 (in a simple example for a scientist with only one paper in some field, 0.1 could represent a paper with other 9 co-authors ($n_p=10$) in a journal indexed in only one field ($m_p=1$); or a paper as solo author ($n_p=1$) in a journal indexed in ten categories ($m_p=10$)). Formally, $P_{sf}(T)$ is defined as:

$$P_{sf}(T) = \begin{cases} 1 & \text{if } X_{sf} > 0.1 \\ 0 & \text{otherwise} \end{cases}$$

We then calculate the number of authors that have participated in fields $f$ and $f'$ before time $T$ by taking the inner product of $P_{sf}(T)$ with itself across all scientists. Formally, we define the matrix $M_{ff'}(T)$ as:

$$M_{ff'}(T) = \sum_s P_{sf}(T) P_{sf'}(T)$$

Finally, we define the proximity between fields $f$ and $f'$ denoted by variable $\phi_{ff'}$ by taking the probability that a scientist with presence in field $f'$ also has presence in field $f$:

$$\phi_{ff'}(\mathrm{T}) = \frac{M_{ff'}}{\sum_s P_{sf'}},$$

where $\sum_s P_{sf'}$ is the total number of scientists that have presence in field $f'$.

$\phi_{ff'}(T)$ is the adjacency matrix representing the research space expressed by the career trajectory of scientists in our dataset observed up to time $T$.

Figure 2 shows a network visualization of the research space ($\phi_{ff'}(2011)$) (i.e. using data from 1971 to 2010). Here nodes are research areas (in UCSD classification) and links connect research areas that are likely to share authors. Colors are assigned according to the main areas defined by the classification, and node sizes are proportional to the total number of papers produced in that area (for papers with multiple categories, we distribute their contribution equally among all of the categories available). Since most proximities are larger than zero, we visualize the network using only the strongest links, which are the links in the Minimum Spanning Tree (MST) and the links for which the conditional probability of sharing authors is larger than 21.2% a threshold that allows to visualize a rich community structure. Furthermore, to simplify the visualization we take only the maximum of the probability between two areas, since the matrix of proximities is not symmetric (a similar visualization of the research space in SCImago classification is provided in the supplementary material).

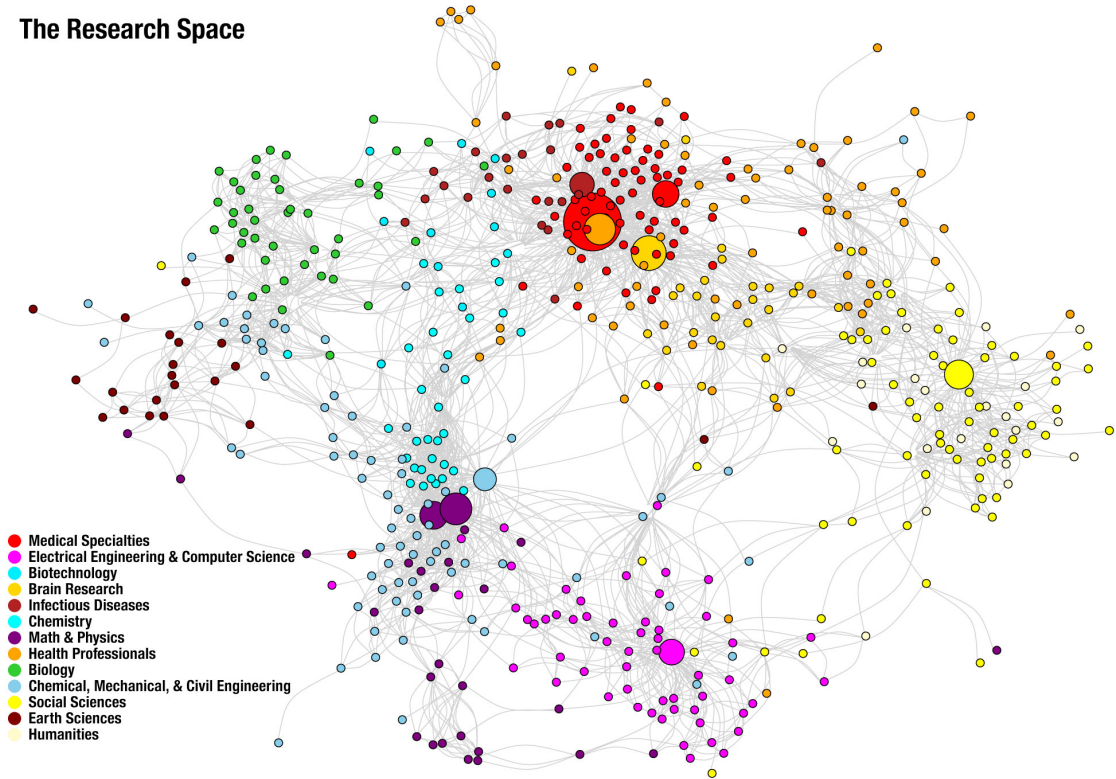

**Figure 2: The Research Space.** Nodes represent research fields and links connect fields that are likely to share authors. The size of nodes is proportional to the number of papers published in that field.

Next, we compare the links in the research space with the UCSD bibliographic coupling science map using a scatter plot and a linear model (Figure 3). Surprisingly, since we expect fields that share authors to cite each other, we find a relatively low correlation ($R^2$ = 0.001) between the links in both maps. For instance, the proximity among "Crustaceans" and "Marine Biology", or "Environmental Protection" and "Water Treatment" in the research space is high, while the volume of citations among both of these pairs of fields in the UCSD science map is low. Conversely, "Cross Disciplinary Studies" with "Ethics", or "Electrochemical Development" and "Metallurgy" are pairs of fields that often cite each other, but share a relatively small number of co-authors. This orthogonally between both maps tells us that predictions made with either of them will likely be dissimilar since the UCSD map is capturing the relatedness or knowledge flows between fields, and the research space is capturing the sharing capacities needed to produce science in different fields.

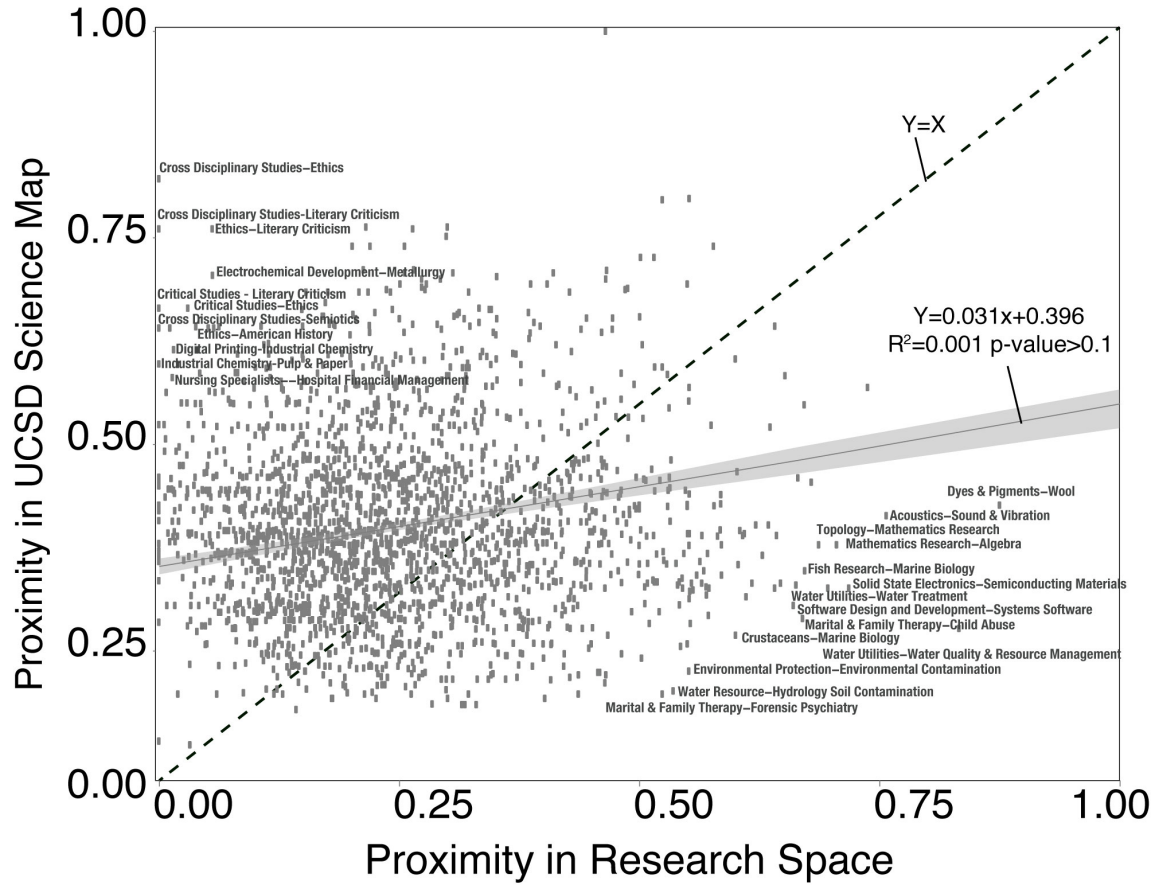

**Figure 3: Comparison between the links estimated for the research space and those reported for the UCSD science map.** Since the matrix of proximities in the research space is by definition not symmetric, we use the maximum value of the link between each pair of areas. The observed low correlation is also true for the minimum, or average.

**Using the research space to predict future research output**

We next use the research space to predict the future presence of an individual, organization, or country in a research field. To make these predictions we define five possible states for individuals, organizations, or countries in a research field. These states are: inactive, active, nascent, intermediate, and developed. To define these states we compare the presence of an individual, organization, or country ($s$), in a research field ($f$), with the presence that we expect from that individual, organization, or country, based on its effective number of papers $X_{sf}$. If the effective number of papers produced by an individual, organization, or country (an entity $s$) in field $f$ is larger than the effective number of papers we expected from an entity with that many total papers in that field, then we say that entity $s$ is developed in the field $f$. Formally we define the level of development of an individual, organization, or country $s$ in field $f$ using the Revealed Comparative Advantage indicator [17] which is defined as:

$$RCA_{sf} = \frac{\frac{X_{sf}}{\sum_f X_{sf}}}{\frac{\sum_s X_{sf}}{\sum_{sf} X_{sf}}}$$

The *RCA* and its normalized version, known in Scientometrics as the Activity Index (AI), have been widely used to analyze the research output of countries [18–22]. Here, we use $RCA_{sf}$ to define the five discrete states that we use to characterize the diversification and evolution of the research output of individuals, organizations, and countries:

| | |
|---|---|
| **Inactive** (with no papers in the field): | $0 = RCA_{sf}$ |
| **Active** (with papers in the field): | $0 < RCA_{sf}$ |
| **Nascent** (with a few papers in the field): | $0 < RCA_{sf} < 0.5$ |
| **Intermediate** (with less papers than expected in the field): | $0.5 \leq RCA_{sf} < 1$ |
| **Developed** (with more papers than expected in the field): | $1 \leq RCA_{sf}$ |

We then predict the probability that individual, organization, or country, $s$ will increase its level of development in field $f$ by creating an indicator of the fraction of fields that are connected to field $f$ and that are already developed by $s$. When we are evaluating transitions to a developed state (to $RCA_{sf}>1$), we define $U_{sf}$ as a matrix that is equal to one when $RCA_{sf} \geq 1$ and 0 otherwise. When we are evaluating the transition from an inactive to an active state (from $RCA_{sf}=0$ to $RCA_{sf}>0$), we define $U_{sf}=1$ when $RCA_{sf} \geq 0$. Using the $U$ matrix we define the density of entity $s$ on field $f$ ($\omega_{sf}$), which is our estimator of the probability that entity $s$ will increase its level of activity in field $f$ as:

$$\omega_{sf} = \frac{\sum_{f'} U_{sf'} \phi_{ff'}}{\sum_{f'} \phi_{ff'}}$$

Finally, to predict a transition of entity in field *f* between a pair of states (i.e. from inactive to active), we look at all fields that are in the initial state (i.e. inactive) and sort them by density ($\omega_{sf}$). The prediction is that the field with higher density will transition to a higher state of development (e.g. from inactive to active), before the fields with lower densities.

For the UCSD science map, we use the same algorithm, but replacing $\phi_{ff'}$ and $\phi_{ff'}$ by the links $\phi_{ff'}$ between fields made available in [3]. The construction of the links of the UCSD science map is detailed in the supplementary material of [3].

## Results

We now use the methodology described above to predict the future presence of an individual, organization, or country, in a field that he or she has not participated in. To measure the accuracy of our predictions we use the area under the Response Operator Characteristics curve (ROC curve). The ROC curve plots the true positive rate of a predictive algorithm (in the y-axis) against its false positive rate (x-axis). A random prediction, having the same rate of true positives and false positives, produces a ROC curve with an area of 0.5, so values between 0.5 and 1 represent the accuracy of the predictive method. The ROC curve is a standard statistic used to measure the accuracy of a predictive method and is related to the Mann-Whitney U-test, which measures the probability that a true positive is ranked above a false positive.

To make our predictions using the research space we construct our proximity matrix using only data from years prior to 2011 (i.e. from 1971 to 2010). We then look at the state (i.e. inactive, active, etc.) of individuals, organizations, and countries for each research field using data from 2008 to 2010 (see examples of overlay maps with the defined states in the supplementary material). Finally, we predict changes in the level of development (i.e. from inactive to active) of each individual, organization, and country, observed between 2011 and 201. In the remainder of the paper we study seven changes in the level of development of an entity in a field. Changes from inactive to active for individuals, institutions and countries, and changes from nascent to developed, and from intermediate to developed for organizations and countries (since RCA values to level of individuals are not meaningful).

Figures 4 a-c, compare the accuracy achieved by the research space and the UCSD science map for the transition from inactive to active. Figures 4 d-e and figures 4 f-g compare the accuracy of the transitions from nascent to developed and from intermediate to developed, respectively. For individuals we only look at transitions from inactive to active, since the nascent and intermediate levels do not make sense for individuals given their limited output (compared to organizations and countries). The distributions of areas under the ROC curve obtained for each transition and method are shown using boxplots (where the horizontal bar is the median, the red circle is the mean, the box contains the interquartile range, and the whiskers encompass more than 96% of the sample). These boxplots describe the distribution for the areas under the ROC curve obtained, respectively, for 4,850 individuals 730 organizations (including research institutions), and 77 countries. The inclusion criteria involved all entities satisfying the inequality

$$\sum_{T=T_0}^{T=T_0+\Delta T} \sum_{f} X_{sf}(T) \geq B\Delta T$$

with B = 3 for individuals, and B = 30 for countries and organizations. The inequality helps us focus on the most productive individuals, organizations, and countries.

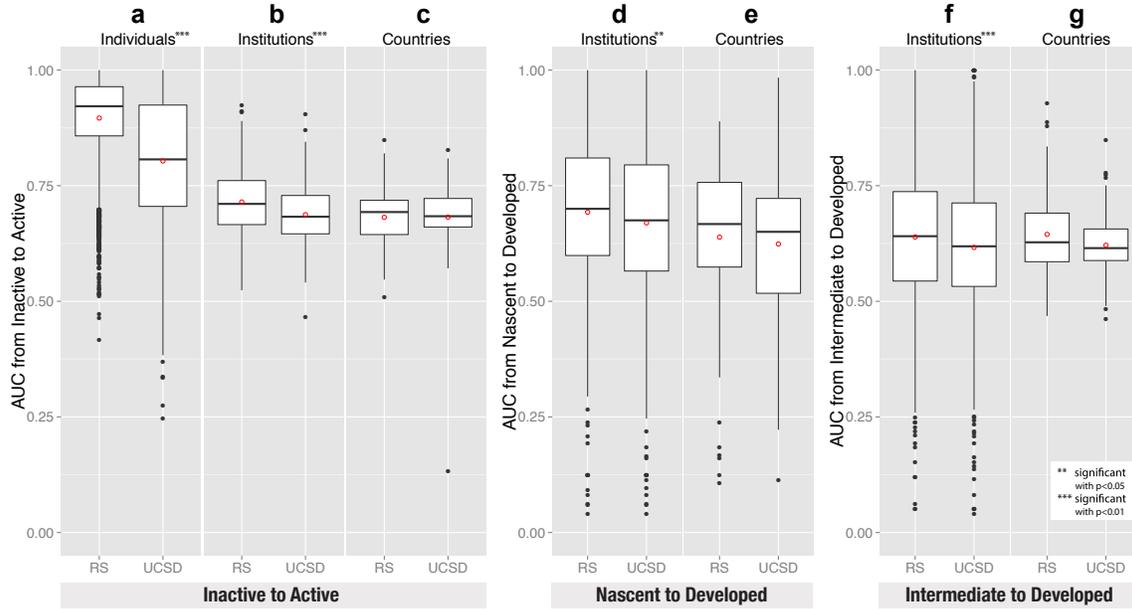

**Figure 4: The predictive power of the Research Space (RS) versus the UCSD science map.** For each entity, a ROC curve is calculated across fields for the given transition. Each boxplot represents the distribution of AUCs. Higher values indicate higher predictive accuracy.

We now focus on transitions from inactive to active (from having no papers in the field ($RCA_{sf}$=0) to having some ($RCA_{sf}$>0)). For both individuals (Figure 3 a) and organizations (Figure 3 b) we find that the predictions made using the research space are significantly more accurate than the predictions made using the UCSD science map. The average area under the ROC curve for individuals (Figure 3 a) is 0.8963 for the research space and 0.8034 for the UCSD science map. This difference is highly statistically significant (ANOVA p-value<0.001). For organizations (Figure 3 b), the averaged accuracy is lower, but the research space is also significantly more accurate than the UCSD science map when it comes to predicting the future presence of a organization in a research field (averages are $AUC_{research\_space}$=0.7148, $AUC_{UCSD\_science\_map}$=0.6873, ANOVA p-value<0.001). For countries, however, both methods are equally accurate (Figure 3 c averages are $AUC_{research\_space}$=0.6816, $AUC_{UCSD\_science\_map}$=0.6819, ANOVA p-value>0.1), indicating that the increase in accuracy observed for the research space expressed itself for more disaggregate units (individuals and organizations).

Now, we focus on transitions from nascent to developed. These are transitions where a country, or organization, went from having a relatively small presence in a research field ($0<RCA_{sf}<0.5$), to a presence that is larger than what is expected from their size and the size of the field ($RCA_{sf}>1$). Once again we find that for organizations (Figure 3 d) the predictions made using the research space are significantly more accurate than the predictions made using the UCSD science map when it comes to predicting the future development of a organization in a research field (averages are $AUC_{research\_space}=0.6927$, $AUC_{UCSD\_science\_map}=0.6696$, ANOVA p-value<0.05). For countries, however, Figure 3 e both methods are equally accurate (averages are $AUC_{research\_space}=0.6387$, $AUC_{UCSD\_science\_map}=0.6239$, ANOVA p-value>0.1), indicating that for transitions from nascent to developed the increase in accuracy observed for the research space is also expressed itself for more disaggregate units (individuals and organizations).

Finally, we look at the transitions from intermediate to developed. These are transitions where a country or organization, went from having a good-sized presence in a research field ($0.5 \leq RCA_{sf}<1$), to a presence that is larger than what is expected from their size and the size of the field ($RCA_{sf} \geq 1$). Once again we find that for organizations (Figure 3 f) the predictions made using the research space are significantly more accurate than the predictions made using the UCSD science map. The average area under the ROC curve for organizations is 0.6390 for the research space and 0.6164 for the UCSD science map. This difference is highly statistically significant (ANOVA p-value<0.01). For countries, however, Figure 3 g both methods are equally accurate (averages are $AUC_{research\_space}=0.6447$, $AUC_{UCSD\_science\_map}=0.6213$, ANOVA p-value>0.05), indicating that for transitions from nascent to developed the increase in accuracy observed for the research space is also expressed itself for more disaggregate units (individuals and organizations).

Table 1 summarizes our results. Rows represent the levels of aggregation (individuals, organizations, and countries), and columns represent the transitions studied (inactive to active, nascent to developed, and intermediate to developed).

| Transition | Inactive to Active ($RCA_i=0$ to $RCA_i>0$) | | Nascent to Developed ($0<RCA_i<0.5$ to $RCA_i\geq 1$) | | Intermediate to Developed ($0.5\leq RCA_i<1$ to $RCA_i\geq 1$) | |
|---|---|---|---|---|---|---|
| Aggregation | Research Space | UCSD Science Map | Research Space | UCSD Science Map | Research Space | UCSD Science Map |
| Individuals | AUC=0.896*** | AUC=0.803 | N/A | N/A | N/A | N/A |
| Organizations | AUC=0.715*** | AUC=0.687 | AUC=0.693** | AUC=0.670 | AUC=0.639*** | AUC=0.616 |
| Countries | AUC=0.682 | AUC=0.682 | AUC=0.639 | AUC=0.624 | AUC=0.645 | AUC=0.621 |

\*\*\* significant with p<0.01 \*\* significant with p<0.05

## Discussion

Understanding the structure of research production is important for scientists, universities, and countries, to understand where they are and where they can go. In this paper we contributed to this literature by introducing the research space, a map of science where links connect pairs of fields if individual are likely to publish in both of them. We used the research space to predict changes in the level of development of individuals, organizations, and countries, for research fields, finding that the research space is a significantly more accurate predictor of the evolution of research output for fine-grained units (individuals and organizations), than the UCSD citation based science map. Both maps, however, are of comparable accuracy when predicting the evolution of the research output of countries, indicating that the research space is particularly relevant for evaluating the research output of individuals and organizations.

Still, there are many questions that this research leaves unanswered. One of these questions is the financial cost required to develop each particular research in an area. Simple intuition tells us that the costs required to develop a field vary enormously for different areas of research. Some research fields require large infrastructure investments, like the advanced facilities needed to perform cutting edge work in biology or the accelerators and reactors needed to make progress on particle or plasma physics. Other areas of research, like data science or economics, can be stimulated by opening more positions for faculty, graduate students, and postdocs, since the infrastructure costs needed to perform research in these fields are modest compared to the ones needed to perform research in more capital intensive fields. In the future, a methodology to evaluate the potential of success of an individual or organization in a field, together with the costs needed to advance research in that direction, would help provide a tool that policy makers could use to strategize the development of research efforts. Our hope is that the methods advanced in this paper are a step in that direction.

## Acknowledgements:

M.G and C.H were supported by the MIT Media Lab Consortia and MIT Chile Seed Fund. M.G was supported by the Universidad de Playa Ancha, Chile (ING01-1516) and the Universidad Técnica Federico Santa María, Chile (PIIC2015). D.H was supported by the Marie Curie International Outgoing Fellowship within the EU 7$^{th}$ Framework Programme for Research and Technical Development: Connecting_EU! - PIOF-GA-2012-328828. M.M was supported by Basal Project FB-0821. C.H was supported by the Metaknowledge Network at the University of Chicago.

## Author Contributions Statement:

C.H wrote the main manuscript text. M.G and C.H prepared all figures. C.H, M.G, M.M and D.H conducted and discussed the experiments. M.A and M.G downloaded, analyzed and curated the data. All authors reviewed the manuscript.

## Competing financial interests
Authors declare no competing financial interests.

# Supplementary Material

# The Research Space: using career paths to predict the evolution of the research output of individuals, institutions, and nations

Miguel R. Guevara[1,2,4], Dominik Hartmann[1,3], Manuel Aristarán[1], Marcelo Mendoza[4], and César A. Hidalgo[1*]

[1]Macro Connections, The MIT Media Lab, Massachusetts Institute of Technology, Cambridge, MA,
[2]Department of Computer Science, Universidad de Playa Ancha, Valparaiso, Chile,
[3]Chair for Economics of Innovation, University of Hohenheim, Germany,
[4]Department of Informatics, Universidad Técnica Federico Santa María, Santiago, Chile

**Content:**

# 1 Analysis of the raw data

The raw dataset consisted of 12,445,334 publications from Google Scholar between 1971 and 2014. After cleaning the dataset for missing data and fake accounts or non-disambiguated data, our final datasets comprised 4,745,774 publications.

Fig. 1 presents the distribution of publications in the raw data, excluding publications with missing information about the publication year. After filtering publications with spurious or inexistent information about the year (i.e. 1900 or 2024) we got 12,293,468 publications in the time period between 1971 and 2014.

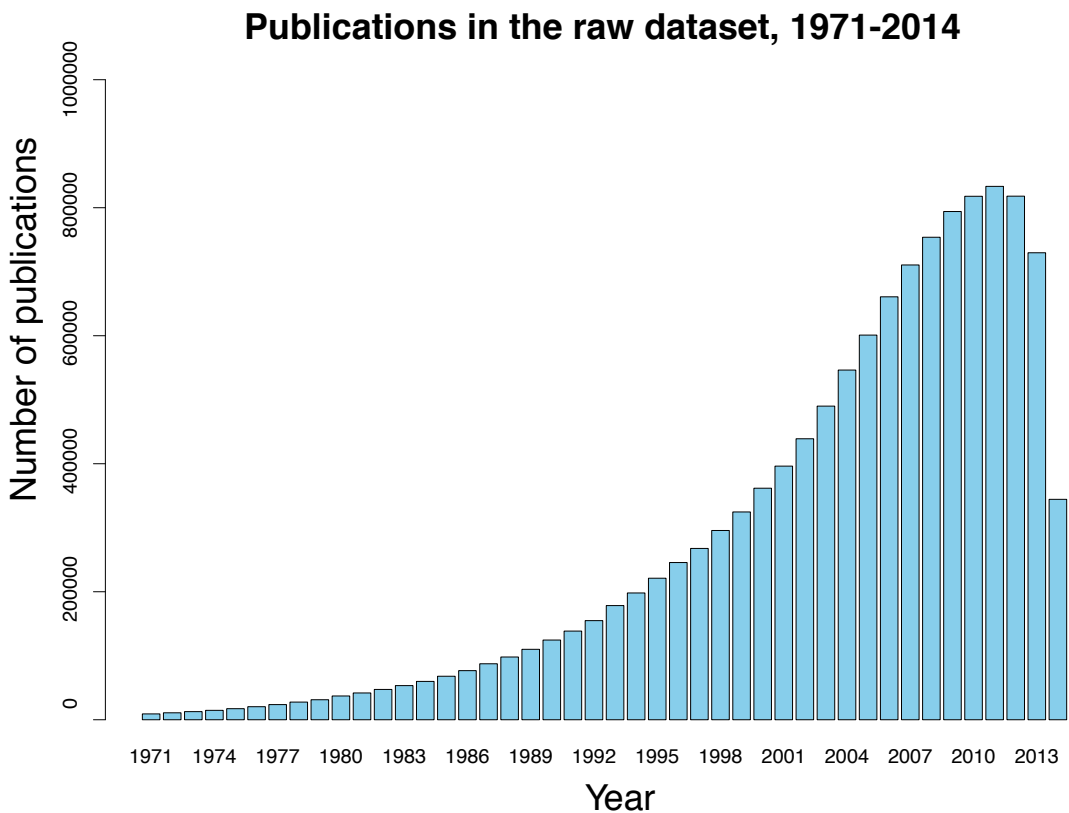

**Fig. 1** Distribution of 12,293,468 publications in the raw dataset. Publications presented in this barplot do not necessarily match with a journal (i.e. they are technical reports or presentations)

Then and in order to find in which areas the scholars are publishing, we matched the text with the name of the journal for each publication in our dataset (Google Scholar) with the text of the name of the journal in the list of journals (provided by the classification, Scimago or UCSD). We only considered publications in which the match was 100%.

Moreover, we filter this dataset by focusing only on scholars with less than fifty publications in each year. Those with more than fifty publications tend to have many publications that are not theirs and are thus miss-assigned.

In Fig. 2 we present the distribution of number of publications per author.

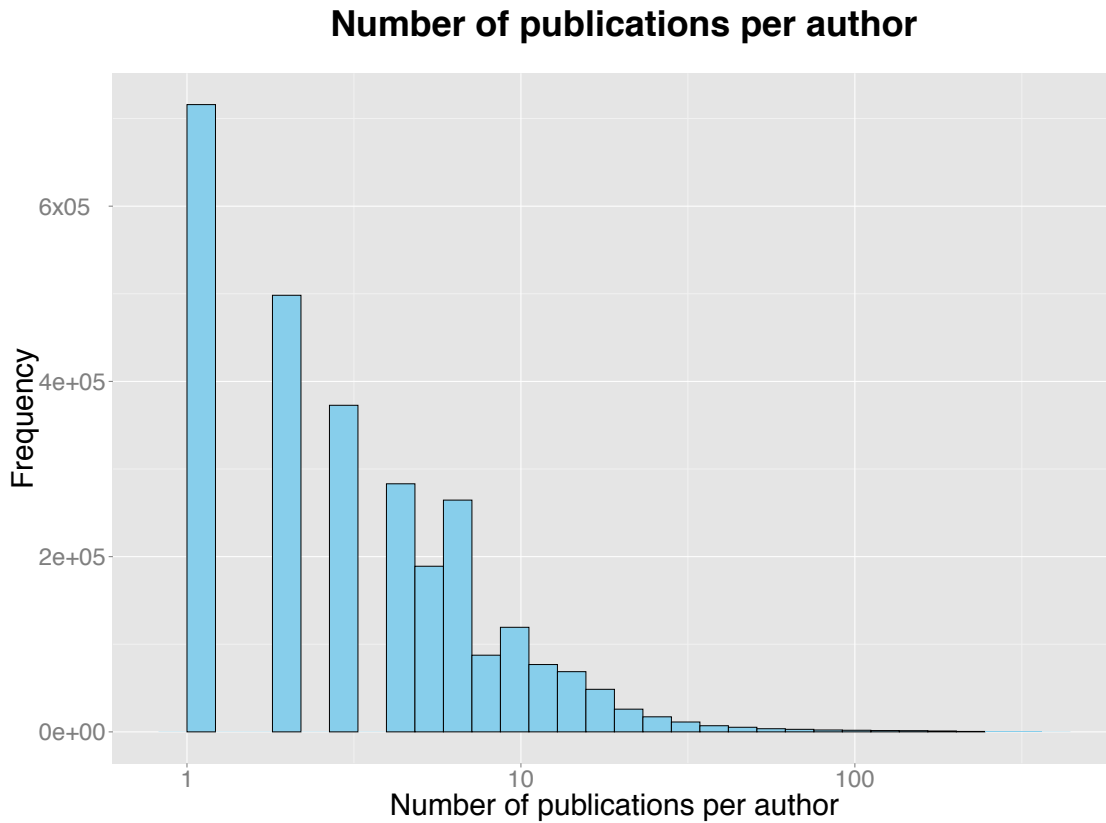

**Fig. 2** Distribution of number of publications per author in each year. Log scale is applied to Y axis

The total number of publications that we used is 4,745,774. The distribution of publications over the years is presented in Fig. 3.

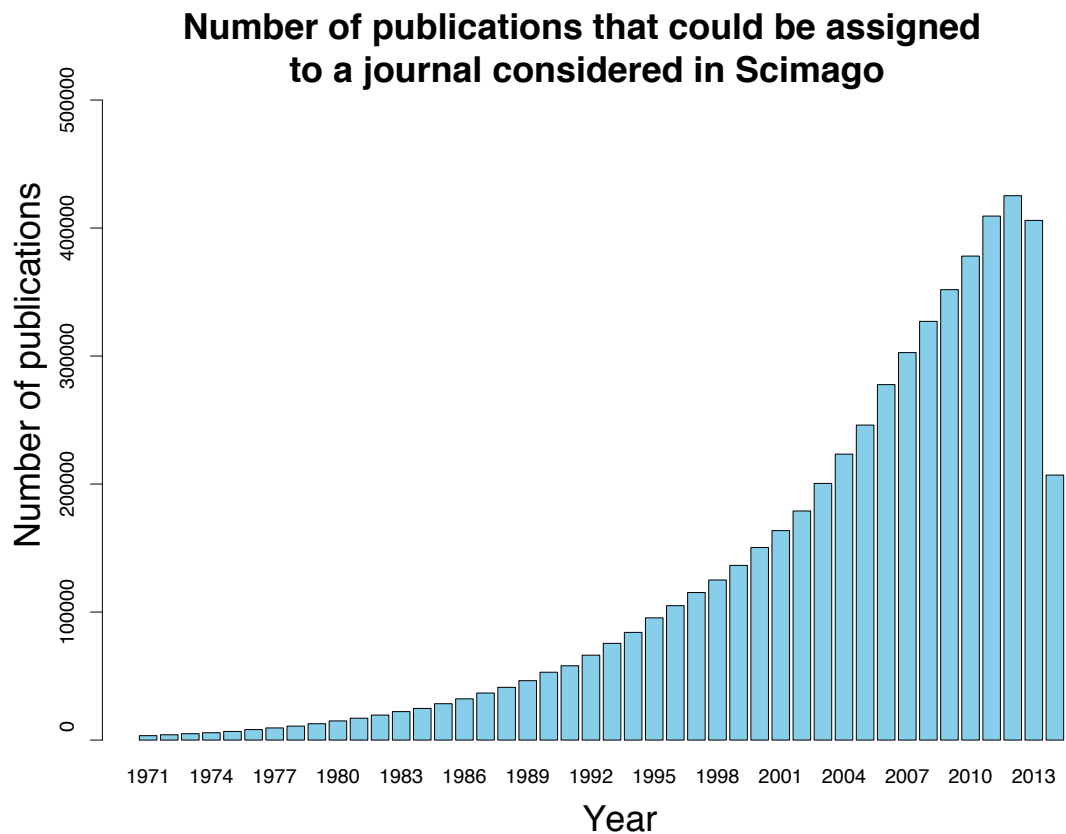

**Fig. 3** Distribution of publications over the years. Publications considered for this plot are publications that belong to a journal in Scimago journal list (see Section 2)

## 2 List of scientific areas and categories

We assign each publication to a research category based on the journal in which it was published using Scopus classification system provided by Scimago that includes 27 main areas of knowledge that are subdivided into 308 fine grained categories.

**Agricultural and Biological Sciences:**
Agricultural and Biological Sciences (miscellaneous); Agronomy and Crop Science; Animal Science and Zoology; Aquatic Science; Ecology, Evolution, Behavior and Systematics; Food Science; Forestry; Horticulture; Insect Science; Plant Science; Soil Science.

**Arts and Humanities:** Archeology (arts and humanities); Arts and Humanities (miscellaneous); Classics; Conservation; History; History and Philosophy of Science; Language and Linguistics; Literature and Literary Theory; Museology; Music; Philosophy; Religious Studies; Visual Arts and Performing Arts.

**Biochemistry, Genetics and Molecular Biology:** Aging; Biochemistry; Biochemistry, Genetics and Molecular Biology (miscellaneous); Biophysics; Biotechnology; Cancer Research; Cell Biology; Clinical Biochemistry; Developmental Biology; Endocrinology; Genetics; Molecular Biology; Molecular Medicine; Physiology; Structural Biology.

**Business, Management and Accounting:** Accounting; Business and International Management; Business, Management and Accounting (miscellaneous); Industrial Relations; Management Information Systems; Management of Technology and Innovation; Marketing; Organizational Behavior and Human Resource Management; Strategy and Management; Tourism, Leisure and Hospitality Management.
**Chemical Engineering:** Bioengineering; Catalysis; Chemical Engineering (miscellaneous); Chemical Health and Safety; Colloid and Surface Chemistry; Filtration and Separation; Fluid Flow and Transfer Processes; Process Chemistry and Technology.

**Chemistry:** Analytical Chemistry; Chemistry (miscellaneous); Electrochemistry; Inorganic Chemistry; Organic Chemistry; Physical and Theoretical Chemistry; Spectroscopy.

**Computer Science:** Artificial Intelligence; Computational Theory and Mathematics; Computer Graphics and Computer-Aided Design; Computer Networks and Communications; Computer Science (miscellaneous); Computer Science Applications; Computer Vision and Pattern Recognition; Hardware and Architecture; Human-Computer Interaction; Information Systems; Signal Processing; Software.

**Decision Sciences:** Decision Sciences (miscellaneous); Information Systems and Management; Management Science and Operations Research; Statistics, Probability and Uncertainty.

**Dentistry:** Dental Assisting; Dental Hygiene; Dentistry (miscellaneous); Oral Surgery; Orthodontics; Periodontics

**Earth and Planetary Sciences:** Atmospheric Science; Computers in Earth Sciences; Earth and Planetary Sciences (miscellaneous); Earth-Surface Processes; Economic Geology; Geochemistry and Petrology; Geology; Geophysics; Geotechnical Engineering and Engineering Geology; Oceanography; Paleontology; Space and Planetary Science; Stratigraphy.

**Economics, Econometrics and Finance:** Economics and Econometrics; Economics, Econometrics and Finance (miscellaneous); Finance.

**Energy:** Energy (miscellaneous); Energy Engineering and Power Technology; Fuel Technology; Nuclear Energy and Engineering; Renewable Energy, Sustainability and the Environment.

**Engineering**: Aerospace Engineering; Architecture; Automotive Engineering; Biomedical Engineering; Building and Construction; Civil and Structural Engineering; Computational Mechanics; Control and Systems Engineering; Electrical and Electronic Engineering; Engineering (miscellaneous); Industrial and Manufacturing Engineering; Mechanical Engineering; Mechanics of Materials; Media Technology; Ocean Engineering; Safety, Risk, Reliability and Quality.

**Environmental Science:** Ecological Modeling; Ecology; Environmental Chemistry; Environmental Engineering; Environmental Science (miscellaneous); Global and Planetary Change; Health, Toxicology and Mutagenesis; Management, Monitoring, Policy and Law; Nature and Landscape Conservation; Pollution; Waste Management and Disposal; Water Science and Technology.

**Health Professions:** Chiropractics; Complementary and Manual Therapy; Emergency Medical Services; Health Information Management; Health Professions (miscellaneous); Medical Assisting and Transcription; Medical Laboratory Technology; Medical Terminology; Occupational Therapy; Optometry; Pharmacy; Physical Therapy, Sports Therapy and Rehabilitation; Podiatry; Radiological and Ultrasound Technology; Respiratory Care; Speech and Hearing.

**Immunology and Microbiology:** Applied Microbiology and Biotechnology; Immunology; Immunology and Microbiology (miscellaneous); Microbiology; Parasitology; Virology.

**Materials Science:** Biomaterials; Ceramics and Composites; Electronic, Optical and Magnetic Materials; Materials Chemistry; Materials Science (miscellaneous); Metals and Alloys; Nanoscience and Nanotechnology; Polymers and Plastics; Surfaces, Coatings and Films.

**Mathematics:** Algebra and Number Theory; Analysis; Applied Mathematics; Computational Mathematics; Control and Optimization; Discrete Mathematics and Combinatorics; Geometry and Topology; Logic; Mathematical Physics; Mathematics (miscellaneous); Modeling and Simulation; Numerical Analysis; Statistics and Probability; Theoretical Computer Science

**Medicine:** Anatomy; Anesthesiology and Pain Medicine; Biochemistry (medical); Cardiology and Cardiovascular Medicine; Complementary and Alternative Medicine; Critical Care and Intensive Care Medicine; Dermatology; Drug Guides; Embryology; Emergency Medicine; Endocrinology, Diabetes and Metabolism; Epidemiology; Family Practice; Gastroenterology; Genetics (clinical); Geriatrics and Gerontology; Health Informatics; Health Policy; Hematology; Hepatology; Histology; Immunology and Allergy; Infectious Diseases; Internal Medicine; Medicine (miscellaneous); Microbiology (medical); Nephrology; Neurology (clinical); Obstetrics and Gynecology; Oncology; Ophthalmology; Orthopedics and Sports Medicine; Otorhinolaryngology; Pathology and Forensic Medicine; Pediatrics, Perinatology and Child Health; Pharmacology (medical); Physiology (medical); Psychiatry and Mental Health; Public Health, Environmental and Occupational Health; Pulmonary and Respiratory Medicine; Radiology, Nuclear Medicine and Imaging; Rehabilitation; Reproductive Medicine; Reviews and References (medical); Rheumatology; Surgery; Transplantation; Urology;

**Multidisciplinary:** Multidisciplinary

**Neuroscience:** Behavioral Neuroscience; Biological Psychiatry; Cellular and Molecular Neuroscience; Cognitive Neuroscience; Developmental Neuroscience; Endocrine and Autonomic Systems; Neurology; Neuroscience (miscellaneous); Sensory Systems.

**Nursing:** Advanced and Specialized Nursing; Assessment and Diagnosis; Care Planning; Community and Home Care; Critical Care Nursing; Emergency Nursing; Fundamentals and Skills; Gerontology; Issues, Ethics and Legal Aspects; Leadership and Management; LPN and LVN; Maternity and Midwifery; Medical and Surgical Nursing; Nurse Assisting; Nursing (miscellaneous); Nutrition and Dietetics; Oncology (nursing); Pathophysiology; Pediatrics; Pharmacology (nursing); Psychiatric Mental Health; Research and Theory; Review and Exam Preparation.

**Pharmacology, Toxicology and Pharmaceutics:** Drug Discovery; Pharmaceutical Science; Pharmacology; Pharmacology, Toxicology and Pharmaceutics (miscellaneous); Toxicology;

**Physics and Astronomy:** Acoustics and Ultrasonics; Astronomy and Astrophysics; Atomic and Molecular Physics, and Optics; Condensed Matter Physics; Instrumentation; Nuclear and High Energy Physics; Physics and Astronomy (miscellaneous); Radiation; Statistical and Nonlinear Physics; Surfaces and Interfaces.

**Psychology:** Applied Psychology; Clinical Psychology; Developmental and Educational Psychology; Experimental and Cognitive Psychology; Neuropsychology and Physiological Psychology; Psychology (miscellaneous); Social Psychology.

**Social Sciences:** Anthropology; Archeology; Communication; Cultural Studies; Demography; Development; Education; Gender Studies; Geography, Planning and Development; Health (social science); Human Factors and Ergonomics; Law; Library and Information Sciences; Life-span and Life-course Studies; Linguistics and Language; Political Science and International Relations; Public Administration; Safety Research; Social Sciences (miscellaneous); Social Work; Sociology and Political Science; Transportation; Urban Studies.

**Veterinary:** Equine; Food Animals; Small Animals; Veterinary (miscellaneous)

## 3 Multiple assignment of journals into categories

The Scimago classification of Science allows multiple indexing of journals into multiple categories. Fig. 4 presents a histogram about the number of categories to which the journals are assigned in Scimago. The maximum number of assignments of a journal in different categories is 12, however, most of the journals are assigned to only one category.

When we measure the presence of a scholar in a category, we normalize her production in a category for a factor $m_p$ which is the number of categories to which the journal is assigned.

See Section **Constructing The Research Space** in the main paper.

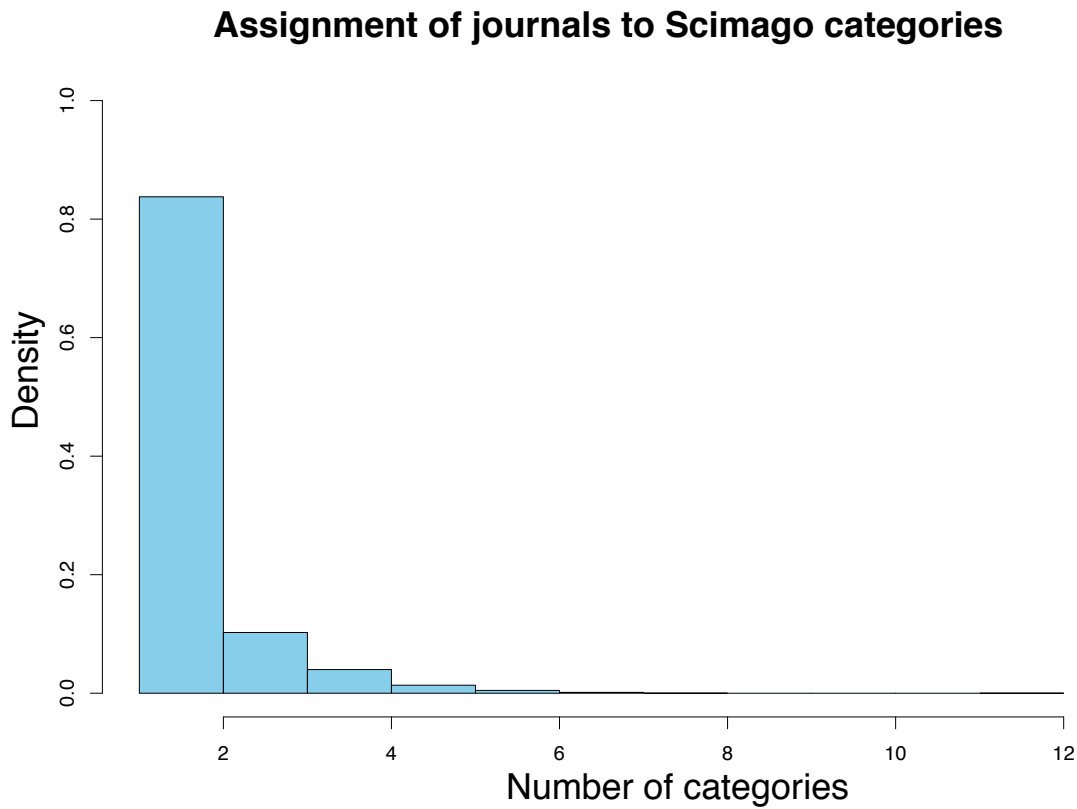

**Fig. 4** Histogram about the number of categories assigned to each journal according to the Scimago classification

## 4 List of areas and categories for UCSD classification

We use the UCSD classification of science in order to perform comparisons between the research space and the UCSD map of science (with the same name) that is a map based on citation patterns. The list includes 554 categories distributed in 13 main areas.

**Biology:** Applied Genetics; Aquaculture; Aquatic Disease; Australian Ecology; Biological Conservation; Botany++; Comparative Animal Physiology; Crop Science; Crustaceans; Ecological Modeling; Ecology; Entomology; Environmental Contamination; Environmental Microbiology; Fish Biology; Fish Research; Forest Science; Freshwater Biology; Genetics; Horticulture; Human Evolution; Insect Physiology; Insects; Mammals; Marine Biology; Marine Pollution; Molecular Biochemical Parasitology; Molecular Biological Evolution; Molecular Ecology; Mycology; Parasitology; Pest Management Science; Plant Disease; Plant Ecology; Plant Physiology; Rangeland Ecology; Sociobiology; Soil Analysis; Weed Management; Wetlands; Wildlife Management; Wildlife Research; Zoology.

**Biotechnology:** BioInformatics; Biotechnology Bioengineering; Biotechnology Trends; Enzyme Microbiological Techniques; Food Engineering; Food Protection; Genomics & Nucleic Acids; Microbiology Biotechnology; Protein Science; Proteomics; Systematics & Evolutionary Microbiology.

**Brain Research:** Affective Disorders; Child & Adolescent Psychiatry; Clinical Neurophysiology; Consciousness; Epilepsy; Forensic Psychiatry; Forensic Science; Geriatric Psychiatry; Geriatrics; Headache; Hearing Research; Magnetic Resonance Imagery; Medical Imaging; Memory & Cognition; NeuroImmunology; Neurology; Neurophysiology & Neuroscience; Neuroscience Methods; Neuroscience; Molecular & Cellular; Neurosurgery; Neurotoxicology; Otolaryngology; Laryngoscope; Physical Therapy; Brain Injury; Psychopharmacology; Psychosis; Schizophia; Sleep; Speech Language & Hearing; Vision.

**Chemical, Mechanical, & Civil Engineering:** Acoustics; Aeronautics & Astronautics; Aerospace; Agricultural Engineering; Alloys; Applied Geophysics; Automotive Engineering; Bulk Solid Handling; Cement & Concrete; Ceramics; Chemical Engineering; Combustion; Composites; Construction; Corrosion; Dams & Tunnels; Defects & Diffusion in Materials; Digital Printing; Dyes & Pigments; Earthquake Engineering; Electrochemical Development; Electrochemistry; Energy Fuel; Environmental Pollution; Environmental Protection; Filtration Membrane; Fluid Engineering; Fluid Mechanics; Fluid Phase Equilibrium; Fractures & Fatigue; Friction Lubrication & Wear; Gas Turbines; Geotechnical Engineering; Heat Transfer; Hydrology Soil Contamination; Industrial Chemistry; Machine Tools; Material Science; Materials Processing; Mechanical Design Engineering; Mechanics of Solids & Structures; Metallurgy; Military Aviation; Mining; Naval Architecture; Nuclear Engineering; Numerical Methods in Engineering; Ocean Coastal Management; Ocean Engineering; Oceanographic Instrumentation; Oil & Natural Gas; Ore Processing; Petroleum Engineering; Printing; Pulp & Paper; Pulp Paper Science; Safety Management; Sensors & Actuators; Soil Quality; Soil Science; Solar & Wind Power; Sound & Vibration; Textile Art; Textiles; Transportation Research; Vehicle System Design; Waste Management; Water Policy; Water Quality & Resource Management; Water Treatment; Water Utilities; Water Waste; Welding; Wood; Wood Components; Wool.

**Chemistry:** Applied Catalysis; Atomic Spectrometry; Carbohydrate Research; Carbon; Catalysis; Chemistry (Russia); Chemistry & Material Science; Chromatography; Electrophoresis; Colloid; Computational Chemistry; Computer Aided Molecular Design; Crystallography; Electro Analytical Chemistry; Environmental Chemistry; EthnoPharmacology; Flavors & Fragrance; Food Chemistry; Green Chemistry; Inorganic Chemistry; Liquid Crystals; Macromolecules & Polymers; Mass Spectrometry; Molecular Physics; Nanotechnology; Organic Chemistry; Paints & Coatings; Pharmaceutical Design; Pharmaceutical Research; Phytochemistry; Surfactants; Thermal Analysis; Toxins.

**Earth Sciences:** Air Quality; Archeological Science; Atmospheric GeoPhysics; Atmospheric Science; Climatology; GeoChemistry; Geodesy; Geographic Information Science; Geology (International); Geology & Tectonics; Geomorphology; GIS (non English); Glaciology; Mineralogy; Oceanography; Paleobiology; Paleogeography; Quaternary Research; Remote Sensing; Sedimentary Geology; Seismology; Water Resource.

**Electrical Engineering & Computer Science:** Antenna; Antennae; Mobile Radio; Applied Optics; Artificial Evolution; Artificial Intelligence; Automatic Control; Broadband Communication; Chip Design & Manufacturing; Circuit Systems; Circuits; Computer Graphics; Computer Modeling and Animation; Computer Networks; Computer Systems Design; Computer Systems Theory; Consumer Electronics; Control Systems; Data Mining; Database Design & Management; Dielectrics; Electrical Networks; Electronic Imaging; Electronics; Fault Tolerant Computing; Functional Programing; Fuzzy Logic; Fuzzy Sets; Hydraulics; Image Processing; Instrumentation; Integrated Circuit Design; Library Science; Information Retrieval; Logic; Machine Learning; Medical Image Processing; Microwaves; Radio Frequencies; Mobile Networks; Neural Networks; Parallel Computing; Pattern Recognition; Photo-Optics; Power Distribution; Power Systems; Power Transmission; Robotic Systems; Robotics; Search Engines; Web Crawling; Security; Cryptography; Signal Processing; Software Design and Development; Solid State Electronics; Speech Recognition; Spyware; Malware; Systems Software; Test Equipment; User Interface Design; Wireless Communication.

**Health Professionals:** Addictive Behavior; AIDS Treatment; Alternative Complementary Medicine; Applied Physiology; Muscle; Arthroscopy; Artificial Organs; Audiology; Behavioral Research Therapy; BioEthics; Biomaterials; Biomechanics; Bone Joint Surgery; Clinical Psychiatry; Dental Education; Dental Research; Drug Discovery; Emergency Medicine; Employee Health Benefit Plans; Forensic Medicine; General Practice; Geriatric Nursing; Gerontology; Hospice Care; Hospital Financial Management; Hospital Management; Hospital Pharmacy; Hypertension; Laser Surgery; Medical Education; Medical Insurance; Medical Libraries; Medical Practice; Medical Records; Medical Screening & Epidemiology; Mental Health Assessment; Mental Health Nursing; Midwifery; Molecular Medicine; Nursing Administration; Nursing Education; Nursing Specialists; Nursing Theory; Nutrition; Obesity; Occupational Health; Optometry; Oral Surgery; Orthodontics; Otolyngology; Head Neck; Pain; Perception Motor Skills; Periodontology; Pharmaco Economics; Physical Therapy; Orthopedic; Plastic Surgery;

Preventive Medicine; Prosthetic Dentistry; Psychiatric Nursing; Psychiatric Services; Psychoanalysis; Public Health; Public Health Service; Public Hospitals; Region & Medical Ethics; Retinal Surgery; Rural Health Care; Spine; Sports Medicine; Substance-abuse Treatment; Trauma.

**Humanities:** American History; Art History; Asian Studies; Biblical Literature; Classics; Contemporary Philosophy; Critical Studies; Cross Disciplinary Studies; English Literature; Ethics; German Studies; Hispanic Studies; Italian Studies; Linguistics; Literary Criticism; Medieval History; Modern Language; Music & Theatre; Opera; Philosophy of Education; Philosophy Psychology; Poetry; Science History; Semiotics; Social History; Socio-Cultural Anthropology.

**Infectious Diseases:** Agricultural Environmental Medicine; Animal Science; AntiMicrobial Agents; Bacteriology; Clinical Microbiology; Cytogenetics & Genome Mapping; Dairy Science; Gene Therapy; Immunology; Molecular Biology Methods; Molecular Cell Biology; Mutation; DNA Repair; Peptides; Poultry Science; Reproduction Veterinary; Sexually Transmitted Diseases; Tropical Medicine; Vaccines; Veterinary Medicine; Veterinary Microbiology; Veterinary Science; Virology; World Health Organization.

**Math & Physics:** Algebra; Applied Math; Astronomy & Astrophysics; Cancer Statistics; Chaos Fractals & Complexity; Computational Math; Design & Analysis of Algorithms; Discrete Applied Mathematics; Functional Analysis; Geophysical Science; High Energy Physics; Mathematical Science (Russia); Mathematics Research; Nonlinear Analysis; Nuclear Instrumentation; Nuclear Physics; Optics & Lasers; Optimization Theory; Photonics; Physics; Current Developments; Plasma Physics; Semiconducting Materials; Simulation; Space Research; Superconductor Science; Surface Coating Technology; Surface Science; Topology.

**Medical Specialties:** AIDS Research; Allergy & Clinical Immunology; Anesthetics & Analgesics; Atherosclerosis; Birth Defects; Bone & Osteoporosis; Cancer (translated); Cardiovascular; Chest & Respiratory; Circulation; Clinical Cancer Research; Clinical Chemistry; Clinical Endocrinology; Clinical Infectious Disease; Clinical Medicine (Romania); Clinical Medicine (translated); Clinical Rehabilitation; Dermatological Surgery; Dermatology; Developmental Biology; Diabetes Care; Diabetes Metabolism; Dietetics; Digestion; Drug Safety; Electrocardiography; Endoscopic Surgery; Endoscopy; Eye; Fertility; Gut; Gynecology Oncology; Heart Failure; Catheters; Hepatology; Hormone Research; Human Molecular Genetics; Impotence; Intensive Care; Kidney; Leukemia; Lung Cancer; Menopause; Molecular Endocrinology; Nuclear Medicine; Obstetrics; Oncology; Ophthalmology; Pathology; Pediatric Research; Pediatrics; Pharmacology Science; Pharmacy; Prenatal Diagnostics; Pulmonary; Radiation Protection; Radiation Therapy; Radiology; Rheumatology; Stem Cells; Surgery; Surgical Oncology; Thoracic & Respiratory; Thoracic Surgery; Thrombosis; Toxicology Applied Pharmacology; Transfusion; Transplantation; Urology; Vascular Surgery.

**Social Sciences:** Agricultural Economics; Applied Economics; BioStatistics; Business Ethics; Child Abuse; Child Development; Communication Research; Computer-Aided Process Planning; Construction & Project Management; Criminology; Decision Support Systems; Developmental Economics; Eating Disorders; Sex Roles; Econometrics; Economics; Education; Education Psychological Measures; Educational Psychology; Engineering Education; Environmental Law; Environmental Management; Environmental Policy; Ethnic Migration; Ethnology; Finance; Financial Accounting; Foreign Policy; GeoPolitics; Higher Education; Human Resource Management; Human Rights; International Conflict; International Development; International Economics; Language Learning; Law; Leadership & Organizational Behavior; Marital & Family Therapy; Marketing; Operations Management; Operations Research; Personality; Political Geography; Political Science; Political Studies; Pragmatics & Discourse; Psychosomatic Medicine; Public Administration; Public Policy; Pyschiatric & Behavioral Genetics; Regional Studies; Reliability Engineering; Research Policy; Technology Management; Rural Studies; School Psychology; Science Education; Social Economics; Social Psychology; Social Work; Sociology; Statistics; Strategic Management; Symbolic Interaction; Teacher Education; Evaluation; Third World Political Economics; Tourism; Urban Studies; Vocational Counseling; World Trade; Law.

# 5 Research Space in Scimago classification

In the paper, we illustrate the research space according to the UCSD classification, here in Figure 11 we illustrate the research space according to the Scimago classification. The UCSD classification includes roughly the double of categories than Scimago classification. While the UCSD classification is obtained using clustering techniques over datasets of Web Of Science and Scopus; Scimago is based entirely in Scopus dataset.

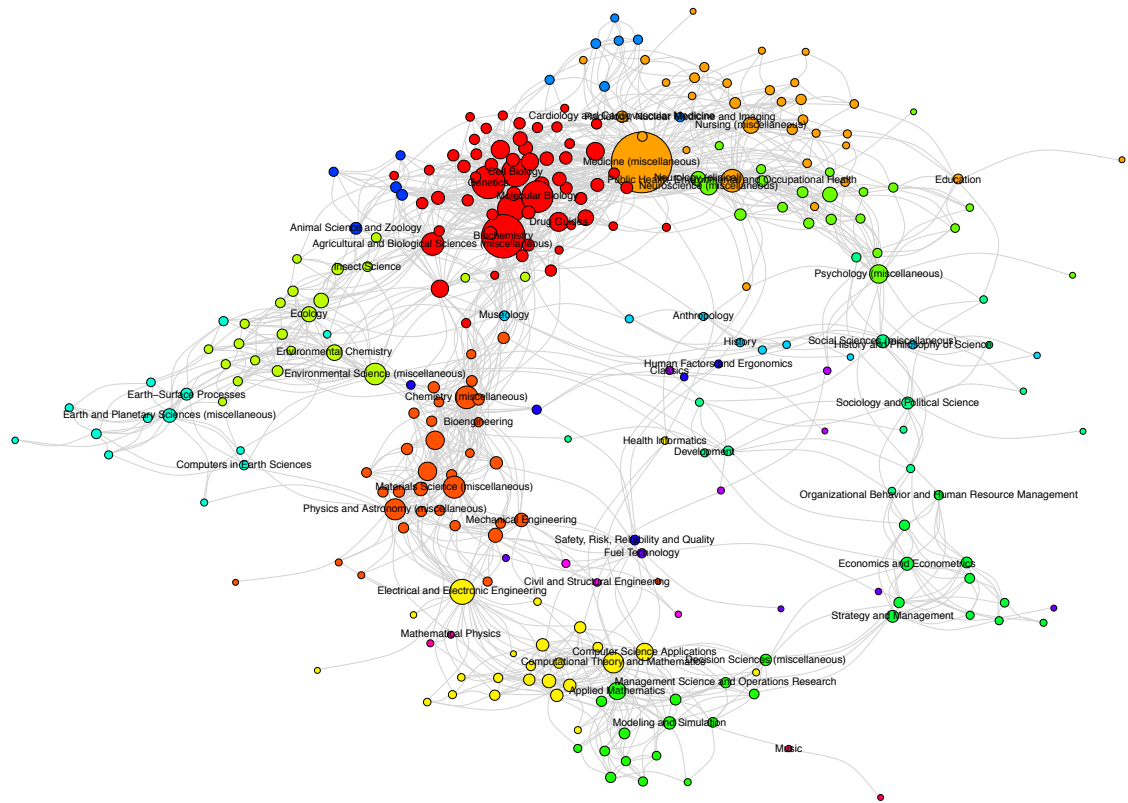

**Fig. 5** The Research Space according to the Scimago classification of Science. Colors of the nodes are defined according to communities detected by using the infomap algorithm. The size of the nodes is proportional to the degree centrality

# 6 Overlay maps for countries and institutions

The research space can be used to visualize the inactive, nascent, intermediate and developed fields of countries, universities / research institutions and scholars (Figures 5-10).

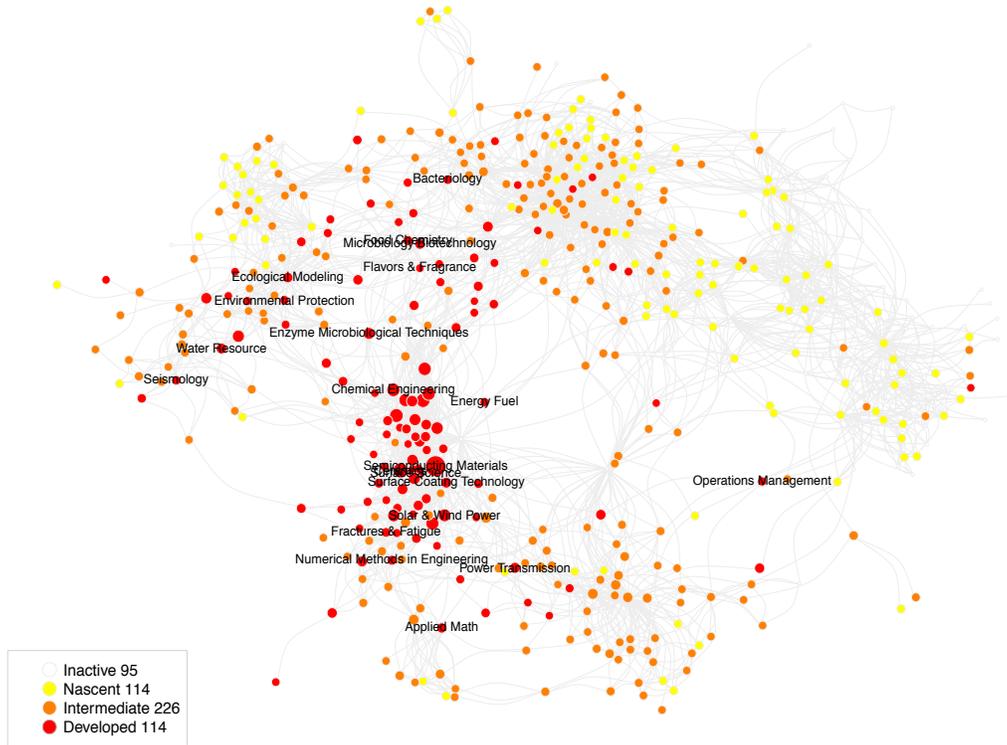

**Fig. 6** Comparative Advantages of India in 2008-2010

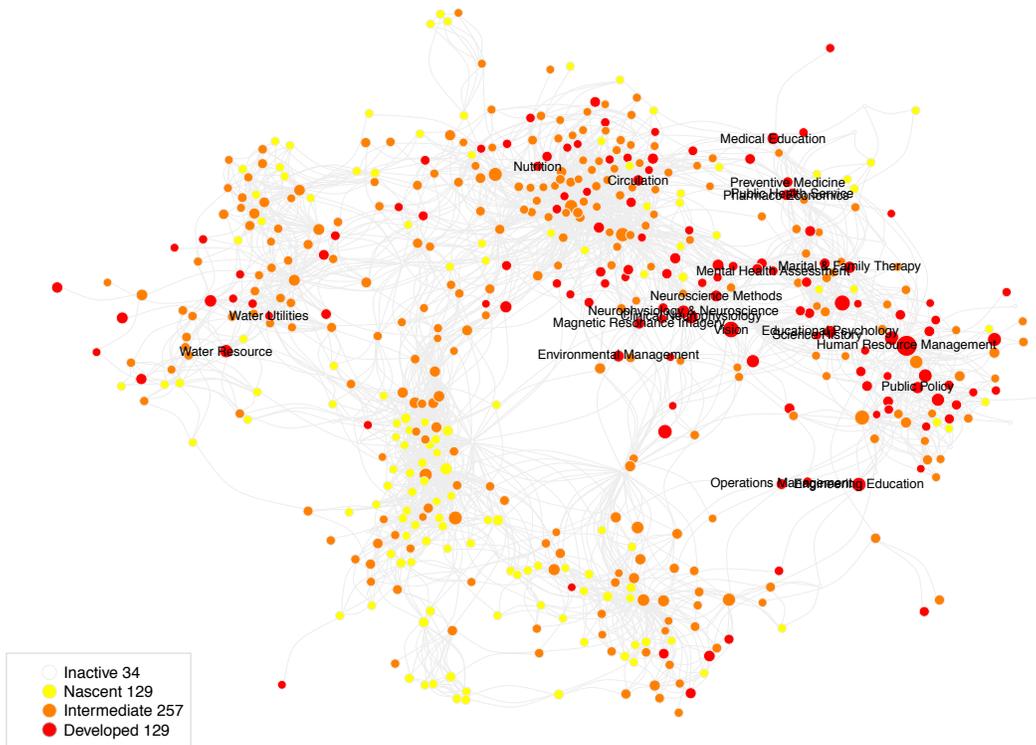

**Fig. 7** Comparative Advantages of Netherlands in 2008-2010

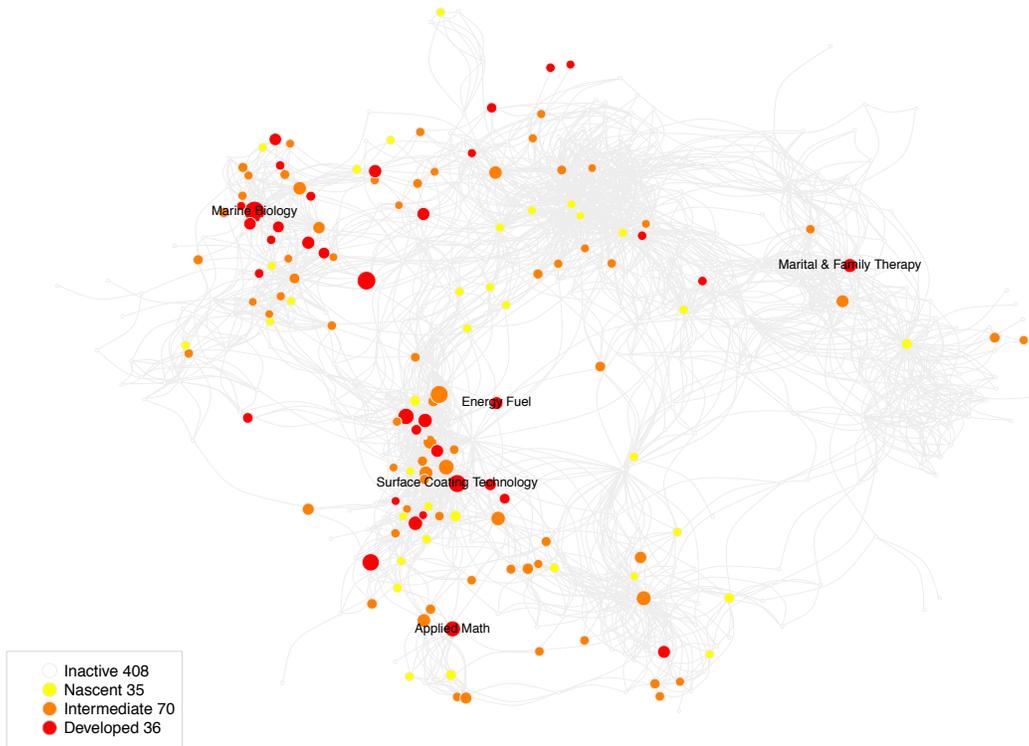

**Fig. 8** Comparative Advantages of Venezuela in 2008-2010